\documentclass[square]{ws-procs11x85}
\usepackage{balance} 

\def\Journal#1#2#3#4{{#1} {\bf #2}, #3 (#4)}
\newcommand{\met}{\hbox{E\kern-0.5em\lower-0.1ex\hbox{/}}_T}

\begin{document}

\twocolumn[
\title{Magnetic Fields in Blazar Pc-scale Jets - Possible
connection to Spin Rates of Black holes ?}

\author{P. Kharb$^1$}
\address{$^1$Dept. of Physics, Purdue University, IN 47907, USA. 
E-mail: pkharb@physics.purdue.edu}
\author{M. L. Lister$^1$, and P. Shastri$^2$}
\address{
$^2$ Indian Institute of Astrophysics, Bangalore, India.} 

\begin{abstract}
We re-examine the differences observed in the pc-scale magnetic field geometry  
of high and low optical polarization Quasars (HPQs, LPRQs) using the MOJAVE 
sample. We find that, as previously reported, HPQ jets exhibit predominantly 
transverse $B$ fields while LPRQ jets tend to display longitudinal $B$ fields. 
We attempt to understand these results along with the different $B$ field 
geometry observed in the low and high energy peaked BL~Lacs (LBLs, HBLs) 
using a simple picture wherein the spinning central black holes in these 
AGNs influence the speed and strength of the jet components (spine, sheath). 
Higher spin rates in HPQs compared to LPRQs and LBLs compared to HBLs could 
explain the different total radio powers, VLBI jet speeds, 
and the observed $B$ field geometry in these AGN classes. 
\end{abstract}

\vskip28pt 
]

\bodymatter

\section{Introduction}
The relativistically beamed active galactic nuclei (AGNs) fall
primarily in two categories on the basis of their total radio power
and optical spectra -- the radio powerful Quasars exhibit strong, broad
and narrow emission lines, while the relatively lower radio power BL~Lacertae
objects display weak or no emission-lines.
In the radio-loud Unified Scheme, FRII and FRI radio galaxies are
considered to be the parent population of Quasars and BL Lacs, 
respectively \cite{UrryPadovani95}.

Quasars and BL Lacs however, do not form a homogeneous class with similar
properties. Low optical polarization radio Quasars (LPRQs) consistently reveal
core optical fractional polarization, $m_{opt}\le3\%$, while the high optical
polarization Quasars (HPQs) routinely reveal highly polarized
optical cores with $m_{opt}\ge3\%$ \cite{AngelStockman80}. The BL~Lacs
likewise seem to fall under two subclasses. The basis for the division is
however their spectral energy distributions (SEDs) -- the synchrotron peaks
of the low-energy peaked BL Lacs (LBLs) lie in the near-IR/optical 
regime, while they lie in the UV/soft X-ray regime for the high-energy 
peaked BL Lacs (HBLs). 

The Quasar and BL~Lac subclasses display additional systematic differences. The 
HPQs and the LBLs seem to exhibit ``extreme'' behaviour in their respective 
classes -- they show greater variability, greater radio core power, greater 
radio core prominence and larger misalignments between their pc- and 
kpc-scale radio jets compared to LPRQs and HBLs, respectively 
(\cite{ListerSmith00} and references in \cite{Kharb07}). 
It has been proposed that these differences are on account of greater 
Doppler beaming in their cores/jets due to being oriented at smaller 
angles to our line of sight. 
Furthermore, Very Long Baseline Interferometry polarization (VLBP) 
observations of the beamed AGNs have revealed that while the majority of 
HPQ jets display transverse magnetic ($B$) fields, 
LPRQ jets display mostly longitudinal $B$ fields \cite{ListerSmith00}. 
While LBLs predominantly display transverse 
$B$ fields in their pc-scale jets \cite{Gabuzda00}, HBL jets tend to 
exhibit longitudinal $B$ fields \cite{Kharb07}.
Orientation alone is insufficient in producing the different $B$ field 
structures in Quasars and BL~Lacs \cite{ListerSmith00,Kharb07}. 

In this paper, we re-examine the trend observed in the pc-scale $B$
field structures of Quasars using a larger sample, and 
attempt to understand the different properties of HPQs, LPRQs, LBLs and
HBLs in the different black hole spin rate scenario.

\section{Objects of Study}
The Quasars belong to the MOJAVE (Monitoring Of Jets in 
AGNs with VLBA Experiments) sample \cite{Lister05}, while the BL~Lacs 
belong to the 1-Jy, HEAO-1 and RGB samples \cite{Kharb07}. 
The classification of Quasars as HPQ/LPRQ was adopted from 
\cite{ImpeyLawrenceTapia91,ListerSmith00}. 
Optical polarization data is however not yet available for all the MOJAVE
sources and additional optical monitoring is necessary.
The VLBP observations of the Quasars and BL~Lacs were made at 15 and 5 GHz, 
respectively. Note that the previous observations that 
showed a difference in the $B$ field structures of HPQs and LPRQs were made at 
22 and 43 GHz \cite{ListerSmith00}.

Quasar jets exhibit pc-scale rotation measures of the order of $\sim$500 
rad~m$^{-2}$ or less \cite{ZavalaTaylor04}. At 15 GHz the expected 
Electric Vector Position Angle (EVPA) rotation is only about 10$^{\circ}$. 
The BL~Lac jets typically exhibit 
lower rotation measures, of the order of a few 100 rad m$^{-2}$ or less. 
Assuming an RM of $\sim$100 rad~m$^{-2}$ would result
in an EVPA rotation of $\sim20^{\circ}$ at 5 GHz.
We therefore do not expect significant changes in the observed $B$ field trend
for Quasars, while it may have some effect on the trend observed in BL~Lacs.

\section{Results}
\begin{figure}[h]
\centering{
\includegraphics[width=7.5cm]{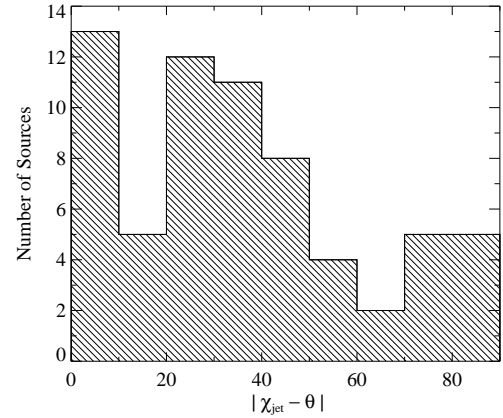}
\includegraphics[width=7.5cm]{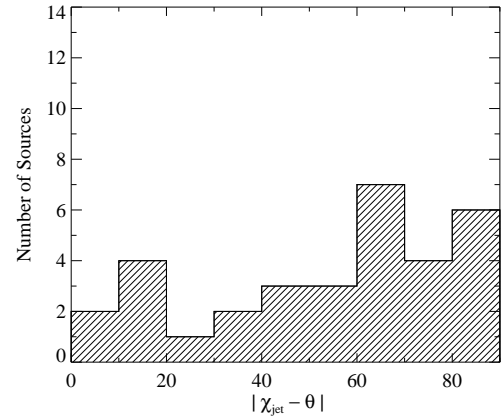}}
\caption{The distribution of the 15~GHz jet EVPA, $\chi_{jet}$, 
{\it w.r.t.} the VLBI jet direction $\theta$ in HPQs (Top) and 
LPRQs (Bottom).}
\label{fig:chi}
\end{figure}

The Quasar sample considered here consists of four times as many HPQs (37) and
twice as many LPRQs (17) than were used by \cite{ListerSmith00}. 
Using this larger sample, 
we find that the trend of different $B$ field structures observed previously,
gains further in significance. The two-sided Kolmogorov-Smirnov (KS)
test indicates that the probability ($p$) of the HPQ and LPRQ EVPA data
being drawn from the same distribution is only 3$\%$ (Fig.~\ref{fig:chi}). 
The probability decreases to 0.3$\%$ when only the inner jet 
(with projected length $r<15$~pc) is considered.

\begin{figure}[h]
\centering{
\includegraphics[width=7.5cm]{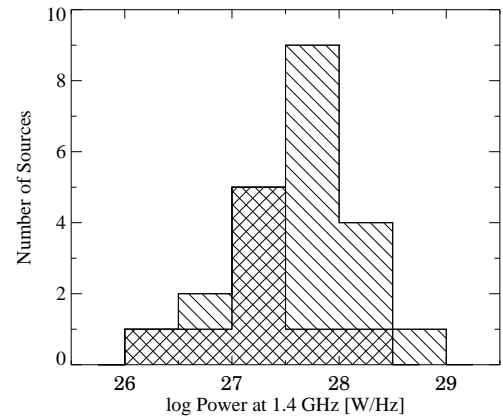}}
\caption{The distribution of total radio luminosity at 1.4 GHz for HPQs (shaded 
with stripes at -45$^{\circ}$) and LPRQs (shaded with stripes at +45$^{\circ}$)
with redshifts $<1.0$.}
\label{fig:power}
\end{figure}

Using the total radio luminosity at 1.4~GHz of the HPQs and LPRQs 
from the NASA/IPAC Extragalactic Database (NED),
we find no significant difference between the HPQ and LPRQ distributions
(KS test $p$ = 55\%). However, when we restrict the redshift range to
$z<1.0$, we find that the HPQs and LPRQs differ with a probability
$>90\%$ (Fig.~\ref{fig:power}). 
The KS test indicated that the redshift distribution of HPQs and LPRQs
was not significantly different before ($p$ = 64\%) or after 
making the cut in redshift ($p$ = 20\%), suggesting that the luminosity
difference in HPQs and LPRQs is not a redshift effect.
However, in the case of the LBLs and HBLs, although the total radio 
power distributions differed significantly ($p$ = 0.001\%,
see \cite{Kharb07}) we could not rule out this bias since the 
redshifts of the LBLs were systematically higher than the HBLs.

Total radio power, however, is affected by Doppler boosting effects in
the core. We are currently reducing Very Large Array (VLA) 1.4 GHz data 
for all the sample objects. This will allow us to directly compare the 
unbeamed extended emission, which will serve as a better indicator 
of jet power, in these sources.


\subsection{Spine-sheath Jets \& Black hole Spins }
Meier\cite{Meier99} has demonstrated through MHD jet simulations
that the jet power can be linked directly to the black hole angular momentum, 
so that jets with greater spins would result in jets with greater speed and 
power. We present here a simple picture by which the different $B$ field 
structures observed in HPQs, LPRQs, LBLs and HBLs, and their different 
total radio powers, can be reconciled. At the core of the argument lies the 
supposition that both FRI and FRII sources have jets with a spine-sheath 
structure, with the faster spine displaying predominantly transverse $B$ fields 
and the slower sheath displaying longitudinal $B$ fields. 
Such a magnetic field configuration could arise due to helical fields
\cite{Laing81}, or due to a combination of wound-up $B$ field lines in the jet
center resulting from a rotating black hole-accretion disk system 
\cite{Meier01}, and a shear layer resulting from jet-medium interaction 
\cite{Laing81}. A sheath could also result due to the flow acceleration 
being a function of the angular distance from the jet axis 
\cite{Ghisellini05}; or on pc-scales, due to an accretion disk wind 
\cite{BlandfordLevinson95}.
Possible spine-sheath $B$ field structures have been observed in both FRI
and FRII jets on kpc \cite{Hardcastle96} and pc-scales 
\cite{Attridge99,Kharb07}. 

Furthermore, we assume that the spin rates of the rotating black holes 
influence both the speed and strength of the spine and sheath,
$i.e.,$ dictates which jet component dominates the overall emission due
to Doppler boosting effects, and the width of the jet.

\begin{figure}
\center
\centerline{\psfig{figure=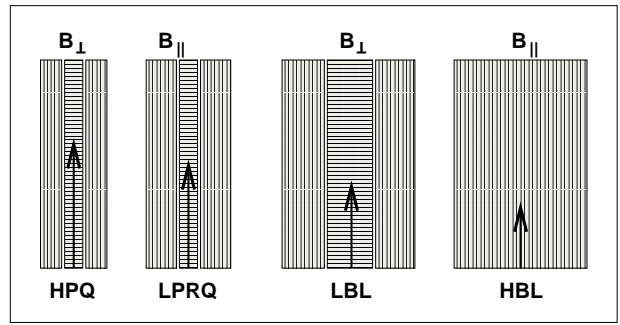,width=8.1truecm}}
\caption{Illustration of a spine-sheath jet structure on pc-scales close to
the central engine. The spine and sheath are shaded with horizontal and 
vertical stripes, respectively, to suggest the dominant transverse and 
longitudinal $B$ fields (note that HBLs could still have a weak 
spine, see text). Spin rates decrease from HPQs to LPRQs and from 
LBLs to HBLs. This influences the jet speed and power, indicated by the
length of the arrows.}
\label{fig:spine}
\end{figure}

Assuming that HPQs and LBLs are intrinsically more powerful than LPRQs and 
HBLs, respectively (Fig.~\ref{fig:power} and \cite{Kharb07}), 
we can interpret the $B$ field geometry in the scenario of different
black hole spin rates. In this picture, the spin rates decrease from the 
HPQs to LPRQs, and from LBLs to HBLs. This is illustrated in 
Fig.~\ref{fig:spine} for a jet with a spine-sheath structure on pc-scales.
Both the spine and sheath contribute to the overall $B$ field
morphology, consistent with the observed EVPAs ranging from 0$^\circ$ to 
90$^\circ$. However, the spine dominates in the highest spin rate HPQs, 
while lower spin rates in LPRQs result in a less dominant spine. 
At small angles to line of sight this would result in predominantly
transverse $B$ fields in HPQs and longitudinal (sheath-dominated) $B$ fields 
in LPRQs. Relatively smaller angles to line of sight are required 
for HPQs to the extent that the narrow spine is not missed. 
Due to the lower spin rates in BL~Lacs, the spine would be weaker and broader.
HBLs may not have a prominent spine at all, resulting in largely 
longitudinal $B$ fields. A weak spine but larger angles to line of sight
in HBLs would also be consistent with the VLBP observations. It is also
possible that the spine is present only on $\mu$-arcsec-scales, as required 
by the TeV data \cite{PinerEdwards04}.

\vspace*{-0.1cm}
\subsection{Jet Speeds, Radio core power \& Misalignment}
\begin{figure}[h]
\centering{
\includegraphics[width=7.5cm]{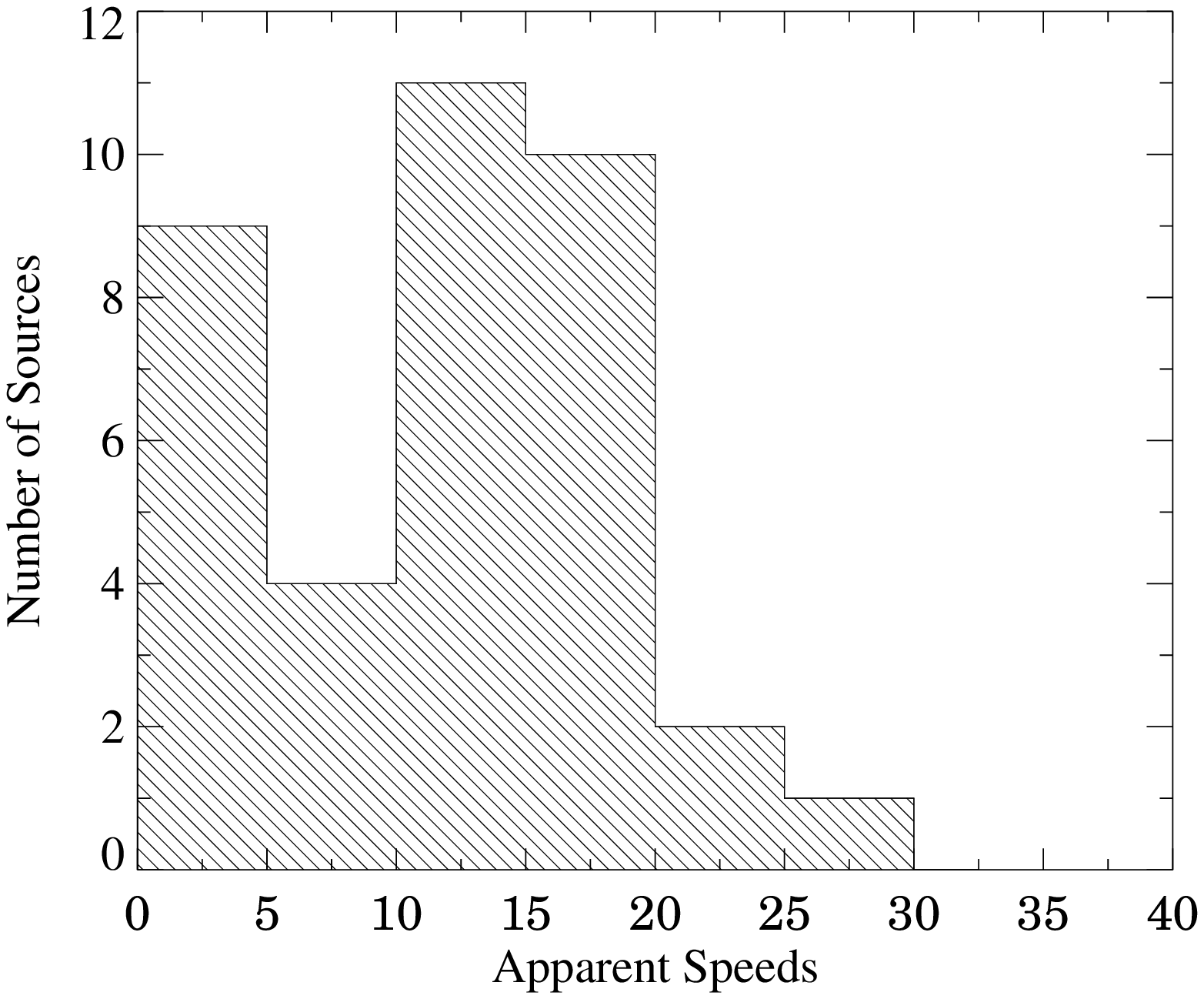}
\includegraphics[width=7.5cm]{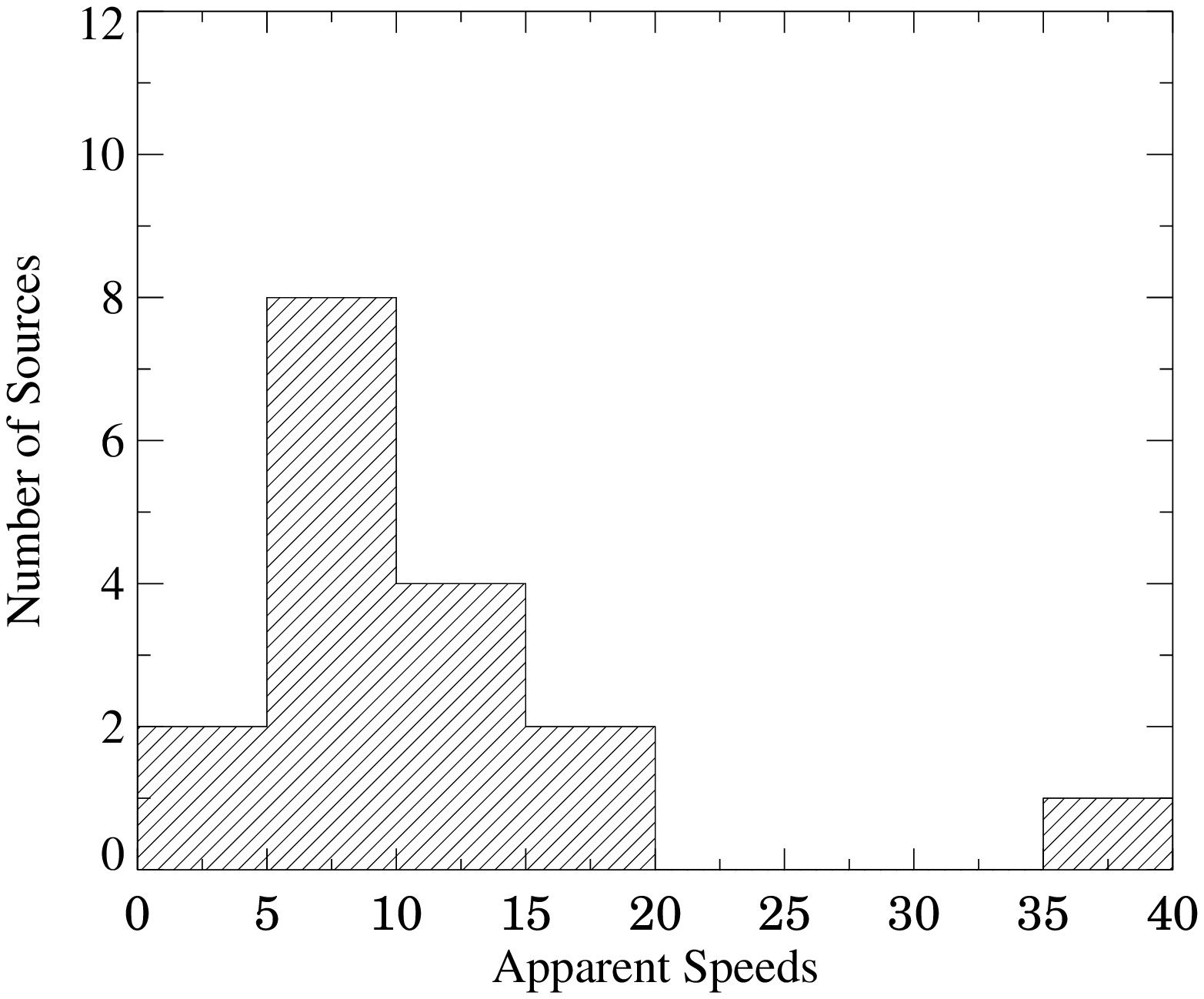}}
\caption{The distribution of jet apparent speeds in HPQs (Top) 
and LPRQs (Bottom).} 
\label{fig:speed}
\end{figure}

We find that the apparent speeds of HPQ jets are systematically higher than 
in the LPRQs (KS test $p>90\%$, Fig.~\ref{fig:speed}; 
Lister {\it et al.,} 2008, in prep.), while LBL jets seem to be faster 
than HBL jets ($p>99.8\%$, \cite{Kharb07}), consistent with the 
different spin rate model. 

As radio cores are the unresolved bases of radio jets, higher radio core
powers in the HPQs and LBLs compared to LPRQs and HBLs, respectively,
are consistent with more powerful radio jets, due to higher black hole 
spins, in these objects. 

Misalignments between pc- and kpc-scale jets can be used as a 
statistical indicator of jet orientation \cite{KapahiSaikia82}.
However, misalignment measurements may be influenced by the
spatial resolution and the dynamic range of the radio images used, and 
must be treated with caution. If HPQs and LBLs indeed have larger 
misalignments between their pc and kpc-scale jets than LPRQs and HBLs, 
respectively, then we could understand that as being due to broader jets 
in LPRQs and HBLs (see \cite{Wiita07} who suggest larger jet opening angles 
in HBLs). In broader outflows, small bends in the jet may not be as apparent, 
leading to seemingly straighter jets and smaller misalignments.
However we note that the misalignment issue is somewhat contentious, with 
different studies reporting different results \cite{Kharb07}. 

\vspace*{-0.1cm}
\subsection{Caveats \& Testing the Model}
The simple model that we have proposed here could not be the complete
picture however. Pc-scale spine-sheath jet structures have not been observed 
in the majority of Quasars or BL Lacs. Although this could be related to the
sensitivity of the VLBP observations, a more complex jet structure is highly
probable. Different black hole spin rates alone cannot account for the 
different emission-line strengths in Quasars and BL Lacs. Differences in the 
nuclear ISM of these AGNs are likely. This in turn will influence the 
spine-sheath structure due to jet interaction.
Differences in the broad-band SEDs of BL Lacs have also been suggested to arise
due to differences in accretion rates \cite{BottcherDermer02}.
A way to test the model would be to examine if the HPQ jets are indeed
the most collimated and narrow, while the HBL jets are the least
collimated and the broadest. Moreover, assuming that black hole spins 
dictate the jet directional stability, misalignments if measurable, should 
progressively increase from HPQs to HBLs.

\section{Conclusions}
We have attempted to understand the different inferred magnetic field 
structures in Quasar and BL Lac subclasses in a simple model wherein the 
black hole spin rates influence the speed and strength of the jet
components (spine, sheath), thereby producing either predominantly 
transverse (dominant spine) or longitudinal (weak spine, dominant sheath) 
$B$ fields. In this model the black hole spin rates decrease from HPQs to 
LPRQs and from LBLs to HBLs. This model is consistent with the 
lower total radio power, weaker radio cores and the lower apparent 
VLBI jet speeds in LPRQs and HBLs, 
compared to HPQs and LBLs, respectively. 
An evolutionary scenario in which LPRQs
evolve from HPQs and HBLs from LBLs, is consistent with the 
proposed model.


\balance


\begin{thebibliography}{}

\bibitem{UrryPadovani95} C. Urry and P. Padovani, \Journal{{\em PASP}}{107}{803}{1995}

\bibitem{AngelStockman80} J. Angel and H. Stockman, \Journal{{\em ARAA}}{18}{321}{1980}

\bibitem{ListerSmith00} M. Lister and P. Smith, \Journal{{\em ApJ}}{541}{66}{2000}

\bibitem{Kharb07} P. {Kharb} {\it et al.}, \Journal{{\em MNRAS}}{384}{230}{2007}

\bibitem{Gabuzda00} D. Gabuzda {\it et al.}, \Journal{{\em  MNRAS}}{319}{1109}{2000}

\bibitem{Lister05} M. Lister and D. Homan, \Journal{{\em AJ}}{130}{1389}{2005}

\bibitem{ImpeyLawrenceTapia91} C. Impey {\it et al.}, \Journal{{\em ApJ}}{375}{46}{1991}

\bibitem{ZavalaTaylor04} R. Zavala and G. Taylor, \Journal{{\em ApJ}}{612}{749}{2004}

\bibitem{Meier99} D. Meier, \Journal{{\em ApJ}}{522}{753}{1999}

\bibitem{Laing81} R. Laing, \Journal{{\em ApJ}}{248}{87}{1981}

\bibitem{Meier01} D. Meier, {\it et al.}, \Journal{{\em Science}}{291}{84}{2001} 

\bibitem{Laing96} R. Laing, \Journal{{\em ASP Conf. Ser. 100: Energy Transport in Radio Galaxies and Quasars}}{100}{241}{1996}

\bibitem{Hardcastle96} M. Hardcastle {\it et al.}, \Journal{{\em MNRAS}}{278}{273}{1996}

\bibitem{Attridge99} J. Attridge {\it et al.}, \Journal{{\em ApJL}}{518}{L87}{1999}

\bibitem{Ghisellini05} G. Ghisellini {\it et al.}, \Journal{{\em A\&A}}{432}{401}{2005}

\bibitem{BlandfordLevinson95} R. Blandford and A. Levinson, \Journal{{\em ApJ}}{441}{79}{1995}

\bibitem{PinerEdwards04} B. Piner and P. Edwards, \Journal{{\em ApJ}}{600}{115}{2004}

\bibitem{KapahiSaikia82} V. Kapahi and D. Saikia, \Journal{{\em JApA}}{3}{465}{1982}

\bibitem{Wiita07} P. Wiita {\it et al.}, \Journal{ArXiv e-prints}{}{0707.3456}{2007}

\bibitem{OrrBrowne82} M. Orr and I. Browne, \Journal{{\em MNRAS}}{200}{1067}{1982}

\bibitem{BottcherDermer02} M. B{\" o}ttcher and C. Dermer, \Journal{{\em ApJ}}{564}{86}{2002}

\end{thebibliography}
\end{document}